\documentstyle[aps,preprint]{revtex}
\begin{document}
\draft
\tightenlines
 
\title{Numerical analysis of the master equation}

\author {Ronald Dickman$^{1}$}

\address{
Departamento de F\'\i sica, ICEx,
Universidade Federal de Minas Gerais, Caixa Postal 702,
30161-970, Belo Horizonte - MG, Brasil}
\date{\today}

\maketitle
\begin{abstract}
Applied to the master equation,
the usual numerical integration methods, such as Runge-Kutta,
become inefficient when the rates associated with various transitions
differ by several orders of magnitude.  We introduce an integration
scheme that remains stable with much larger time increments
than can be used in standard methods.
When only the stationary distribution is required, a direct iteration
method is even more rapid; this method may be extended
to construct the {\it quasi-stationary} distribution of a process with an
absorbing state.  Applications to birth-and-death processes reveal
gains in efficiency of two or more orders of magnitude.
\vspace {0.6cm}

\end{abstract}


\newpage

The master equation is the basic tool for describing Markovian stochastic
processes on a discrete state space, in continuous time \cite{vanK,gardiner}.  Despite its
central role in stochastic analysis in the physical and biological sciences,
only a handful of exactly soluble examples are known.  Thus the need to integrate 
the master equation numerically arises frequently, and with it the issue of 
computational efficiency.

Consider a Markov process with transition rates $W_{n,m}$ from
state $m$ to state $n$.  In the master equation (ME),

\begin{equation}
\dot{p}_n = \sum_{n'} W_{n,n'} p_{n'} - p_n\sum_{n'} W_{n',n}  \;\;\; (n=0,...,N),
\label{me}
\end{equation}
the factor $\sum_{n'} W_{n',n} \equiv w_n$ multiplying $p_n$ can become large in 
a typical birth-and-death
process, where $W_{n',n} \propto n$ or some higher power of $n$.
This poses a problem for numerical integration via the usual
discretization schemes, such as the Runge-Kutta method (RKM). Since instabilities appear
when $|w_n \Delta t | \geq 1$, we must use a small time increment, 
$\Delta t \sim 1/(\max_n w_n)$,
and convergence is slow.   With some (but not all) of the transition rates
large, the ME is, in effect, a {\it stiff} system of differential
equations, requiring special numerical treatment \cite{numrec}.
In this note I introduce numerical integration schemes for the
ME that are efficient when the transition rates vary over a wide range,
and an iteration method that eliminates the need for step-by-step
integration, when only the stationary (or quasi-stationary)
distribution is required.   Simple but detailed examples are used to
illustrate the methods; further applications will be reported elsewhere
\cite{rdunp}. 
(We consider {\it stationary} Markov processes, i.e., time-independent
rates, although the method is not limited to this class of problem.)

We begin by writing the ME in the form

\begin{equation}
\dot{p}_n = -w_n p_n + r_n \;,
\label{me1}
\end{equation}
where 

\begin{equation}
r_n (t) = \sum_{n'} W_{n,n'} p_{n'}(t) \;. 
\label{defr}
\end{equation}
Note that $p_n$ does not appear
in the sum for $r_n$, since $W_{n,n} \equiv 0$.  Both $w_n$ and $r_n$
are nonnegative. (In fact $w_n$ is zero only if state $n$ is
absorbing.)
Integrating Eq. (\ref{me1}) we have

\begin{equation}
p_n(t) = e^{-w_n t} p_n(0) + \int_0^t dt' e^{-w_n(t-t')} r_n (t') \;.
\label{fs}
\end{equation}
This is only a formal solution, since we need the
$p_n (t)$ to evaluate $r_n$, but it is a useful starting point for 
approximations.
If we adopt a time increment $\Delta t$ such that $r_n (t)$ is
approximately constant over this interval, then we have

\begin{equation}
p_n(\Delta t) \simeq e^{-w_n \Delta t} p_n(0) + 
\left[ 1-e^{-w_n \Delta t} \right]
\frac{ r_n (0)}{w_n} \;.
\label{its}
\end{equation}
This relation can then be iterated: we use $p_n (\Delta t)$ to
evaluate $r_n$ at the start of the second interval, and thereby find
$p_n (2\Delta t)$, and so on.  This simple formal integration (FI) scheme, analogous to 
Euler's method for direct numerical integration, already represents
a significant advantage over the usual approaches
when some of the
$w_n$ are large.  The reason is that the exponential factor is
already included in the solution, whereas in the usual discretizion
methods it has to be constructed term by term, in powers of
$w_n \Delta t$.  

Suppose that the ME of interest possesses a unique, stable
stationary solution $\overline{p}_n$.  For a stationary solution,
$\overline{p}_n = \overline{r}_n/w_n$, where $\overline{r}_n$
is given by Eq. (\ref{defr}) with $p_n = \overline{p}_n$.  Setting
$p_n = \overline{p}_n$ on the r.h.s. of Eq. (\ref{its}) we immediately
see that $\overline{p}_n$ is a stationary solution of our iteration
method.  To investigate the linear stability of the stationary
solution, let $p_n = \overline{p}_n + \delta_n$.
According to the ME,

\begin{equation}
\dot{\delta}_n = \sum_{n'} W_{n,n'} \delta_{n'} - \delta_n \sum_{n'} W_{n',n} \;,
\end{equation}
or, in matrix notation,

\begin{equation}
\frac{d}{dt} \delta = S \delta
\label{evdel}
\end{equation}
where 
$S_{n,m} = W_{n,m}$ for $n \neq m$ and $S_{n,n} = -w_n$.
Since, by hypothesis, the stationary solution 
$\delta = 0$ is stable, $S$
has one zero eigenvalue and all others negative.

If we insert $p_n = \overline{p}_n + \delta_n$ in the r.h.s. of
Eq. (\ref{its}) (so that $r_n = \overline{r}_n + \sum_{n'} W_{n,n'} \delta_{n'}$),
we find

\begin{equation}
\delta (t+\Delta t) = T \delta (t)
\end{equation}
where $T_{n,m} = (1-e^{-w_n\Delta t}) W_{n,m}/w_n$ for $n \neq m$,
and $T_{n,n} = e^{-w_n\Delta t} $.
Expanding to first order in $\Delta t$, we see that Eq. (\ref{evdel})
applies in this case as well, so the stationary solution is
stable if it is so for the original ME.

We refer to the integration method embodied in relation Eq. (\ref{its})
as a ``first-order FI scheme" since the error in the solution $p_n (t)$
is proportional (for fixed $t$) to $\Delta t$.  [In each step, the error
incurred in treating $r_n$ as constant is ${\cal O}(\Delta t)^2$
(since in fact $r_n(\Delta t) = r_n(0) +  r'_n(0) \Delta t+ \cdots$),
and we require $N = t/\Delta t $ steps.]
An obvious way to improve accuracy, analogous to going from Euler's
method to the midpoint (or second-order Runge-Kutta) method, is to 
replace the assumption of a constant $r_n$ on the interval
$[t,t+\Delta t]$ with a linear approximation.  To do this we first use
Eq. (\ref{its}) to find $p_n (t+\Delta t)$, and then estimate the
$r_n (t+\Delta t)$ using these values.  
Then we form the linear approximation

\begin{equation}
r_n(s) \simeq r_n(t) +  (s\!-\!t) r'_n
\;, \;\;\;\;\; (t \leq s \leq t + \Delta t) \;,
\label{its1}
\end{equation}
where

\begin{equation}
r'_n =  \frac{r_n(t+\Delta t) - r_n (t)}{\Delta t}
\end{equation}

Inserting this in the formal solution, Eq. (\ref{fs}), we find

\begin{equation}
p_n(t+\Delta t) \simeq e^{-w_n \Delta t} p_n(t) 
+ \left[1-e^{-w_n \Delta t} \right] \frac{ r_n (t)}{w_n} 
+ \left[\Delta t - \frac{1-e^{-w_n \Delta t}}{w_n} \right]
 \frac{r'_n}{w_n}
\label{its2}
\end{equation}
(Note that we do not use the expression for 
$dr_n/dt$ that follows from the master equation
in place of $r'_n$, since
it does not, in general, represent the behavior of $r_n$ over a substantial
time interval.)
The total error associated with the second-order FI procedure is $\propto (\Delta t)^2$.
Further improvement is clearly possible, by introducing higher order polynomial
approximations to the $r_n$, but the second-order scheme is quite adequate for the
applications considered here.  The error estimates apply
for a fixed, finite time; the error for $t \to \infty$ is
{\it zero}, if the system possesses a unique stationary distribution.

A tremendous simplification and speedup is possible if
one only requires the {\it stationary} probability distribution.
The relation 
$\overline{p}_n = \overline{r}_n/w_n$ suggests an iterative
procedure of the form

\begin{equation}
p'_n = a p_n + (1-a) \frac{r_n}{w_n}
\label{its3}
\end{equation}
where $0 < a < 1$ is a parameter.  We expect $p'_n$ to converge to
$A \overline{p}_n$, where the factor $A$ depends on $a$ and on the
initial distribution.

Some insight into the choice of
$a$ is afforded by the simple example of a two-state system with
transition rates $W_{1,0} = \lambda$ and $W_{0,1} = \mu$.
The evolution of the probability distribution is given
(in matrix notation) by $p' = M p$, where

\begin{equation}
M= 
\left( \begin{array} {cc}
a & (1-a)\frac{\mu}{\lambda} \\
  &                           \\
 (1-a)\frac{\lambda}{\mu} & a \\
        \end{array} \right) 
\end{equation}
Matrix $M$ has eigenvalues 1 and $2a-1$.  Thus the
iterative procedure converges to a multiple of the stationary distribution, 
$(\mu, \lambda)$ for $0 < a < 1$, and is instantaneous for $a=1/2$.  

A similar analysis of the three-state processes ($n=0,1,2$), with rates

a) $W_{n+1,n} = \gamma$, $W_{n-1,n} = 1$,

\noindent and 

b) $W_{n+1,n} = \gamma$, $W_{2,0} = \gamma^2$, and 
$W_{n,m}=1$ for $n<m$, 

\noindent yields $a=1/3$ as the optimal choice.
(By optimal choice we mean the value that minimizes 
$\max_i |\omega_i |$, the $\omega_i$ being the eigenvalues
of $M$, excluding, of course $\omega_0 \equiv 1$, associated 
with the stationary distribution.)
These results suggest that for processes with a large number of states,
even smaller values of $a$ may be advantageous.  The numerical examples
discussed below support this notion.

A simple modification of the iterative scheme, Eq.~(\ref{its3}),
generates the {\it quasi-stationary} distribution (if such exists)
of a Markov processs with an absorbing state.  Consider a process
on the states $n = 0, 1, 2,...N$ with $n=0$ absorbing, i.e.,
$W_{n,0} \equiv 0$ for all $n$, while $W_{0,n}$ for at least one
$n > 0$.  
(It seems reasonable to exclude manifestly transient states,
in other words, we assume that for each $n$, $W_{n,m}> 0$ for at least
one $m$.)
In this case the stationary state is 
$\overline{p}_n = \delta_{n,0}$, but it is possible that the 
probability distribution, 
{\it conditioned on survival}, 
attains a stationary form, that is, for long times
$p_n (t) \to C(t) q_n$ ($n > 0$), where the $q_n$ are time-independent.
Such quasi-stationary distributions arise in birth-and-death  
processes with saturation, for example the Malthus-Verhulst process
or the contact process \cite{rdunp}.

The defining feature of the quasi-stationary distribution is that 
relative probabilities $p_n (t)/p_m (t)$ ($n$, $m > 0$) are constant,
or equivalently,
$\dot{p}_n/p_n = \kappa $, 
constant and independent of $n$ for $n \geq 1$.
Suppose the probability distribution has attained a quasi-stationary 
form at some time $t$, and that at this instant the distribution
is normalized so: $\sum_{n \geq 1} p_n(t) = 1$. Using Eq. (\ref{me1}) 
we have $r_n = (\kappa + w_n) p_n$, and summing on $n \geq 1$ yields

\begin{equation}
\kappa = \sum_{n \geq 1} \dot{p}_n = - \sum_{n\geq 1} W_{0,n} p_n 
\equiv  -r_0 \;.
\label{r0}
\end{equation}
($r_0$ is the decay rate of the survival probability in the 
quasi-stationary regime.)
Thus we find that in the quasi-stationary state,

\begin{equation}
p_n = \frac{r_n}{w_n - r_0} \;, \;\;\;\; (n \geq 1).
\label{qss}
\end{equation}
This relation suggests that we iterate as follows:

\begin{equation}
p'_n = a p_n + (1\!-\!a) \frac{r_n}{w_n - r_0}  \;.
\label{itqss}
\end{equation}
The new distribution $p'$ should be normalized
after each iteration, since Eq. (\ref{qss}) assumes this
property.
[If $r_0 = 0$ the process of course possesses a true stationary
distribution and Eq. (\ref{itqss}) reduces to Eq. (\ref{its3}).]
We have verified that this scheme converges to the quasi-stationary
distribution much more rapidly than via integration of the ME.

As a first example we consider the nonlinear one-step process
(on  states $n = 1, 2, 3, ...$), with nonzero transition rates:

\begin{equation}
W_{n-1,n} = n(n\!-\!1) \;,  \;\;\;\;\; W_{n+1,n} = \lambda
\label{rtpa}
\end{equation}
This represents the coagulation process $A + A \to A $ with
addition of species $A$ at rate $\lambda$, in a well-stirred system.
In Fig. 1 we compare the mean population size $\langle n \rangle_t$
as furnished by integration of the ME via a fourth-order RKM
(solid line) with the second-order FI scheme,
Eq. (\ref{its2}), (open circles).  
[Here $\lambda = 100$, $p_n(0) = \delta_{n,1}$,
and we have set $p_n(t) \equiv 0$ for $n>100$.]
The results are nearly identical, but the RKM
requires and integration step of $\Delta t \leq 2 \times 10^{-4}$
for stability, whereas the FI integration uses $\Delta t = 0.002$.
The latter yields $\langle n \rangle_t $ with a relative error of less than
$ 0.8 \%$, the largest deviation occurring at relatively short times.
(As noted above, the stationary
values are identical in all cases.)  To maintain
fidelity to the solution at short, as well as long times, 
one may use the FI with an adjustable time step ($+$ in Fig. 1),
such that $r'^* \Delta t $ is small, where $r'^* = \max_n \{r'_n\}$.  
For example, using the second-order FI with a time step of 
$\Delta t = 0.0005 + .02/(1 + r'^*)$, the relative error is 
reduced to $ \leq 0.06 \%$, while the average time step over the region
of interest is 0.0024.

Fig. 2 shows the stationary probability distribution for 
$\lambda = 400$.  The solid line represents the RKM result, while the
points come from iteration of Eq. (\ref{its3}).
The distributions are identical to within one part in $10^5$.
In this case, the distribution converges (such that the error in
$\langle n \rangle_t$ is $< 10^{-10}$), after about 900 iterations,
when we use $a=1/2$.  As $a$ is reduced, the number of iterations
required falls steadily; in fact the procedure converges even for $a=0$,
after only 460 steps.

Our second example is a multi-step generalization of the Malthus-Verhulst
process, with transition rates:

\begin{equation}
W_{m,n} = 
\left\{ \begin{array} {lc}
n [1 \!+\!\mu (n\!-\!1)] e^{\gamma(m-n+1)} \;, & \;\;\;\; m < n \\
\lambda n e^{\gamma(n-m+1)} \;, & \;\;\;\; m > n           \\
0 \;, & \;\;\;\; m=n \\
\end{array} \right.
\label{gmvp}
\end{equation}
for $n$, $m = 0, 1, 2,...$. 
(Thus transition rates between states $n$ and $m$ fall off
$\propto \exp[-\gamma |m-n|]$.  We choose this example to show that
the methods proposed here are not limited to one-step processes,
which admit a relatively simple analysis \cite{vanK,rdunp}.)
In this case $n=0$ is absorbing.  Figure 3
compares the quasi-stationary distribution 
for $\lambda = 3$, $\mu = 0.1$, and $\gamma = 1$, obtained via the RKM
(with $\Delta t = 10^{-4}$; larger values produce instability), 
with iteration of Eq. (\ref{itqss}).
As is clear from the plot, the results are identical in every detail.
The RKM requires about 1270 seconds of cpu on a DEC-alpha workstation
to converge to the quasi-stationary distribution;  
the FI method requires about 50 sec.
Iteration of Eq. (\ref{itqss}), (with $a=0.5$), by contrast,
converges in 9 seconds (about 1600 steps).  
Reducing parameter $a$ to 0.1 yields convergence
in just 900 steps (5 sec. cpu), and in 815 steps for $a=0$.

The steady decrease in computation time as we reduce $a$ leads one
to ask whether it is possible to use $a$ {\it negative}.
Iteration of Eq. (\ref{itqss}) with $a<0$ does not work, as
negative probabilities are generated, but we can exclude these
by writing

\begin{equation}
p'_n = \max \left[0, \; a p_n + (1\!-\!a) \frac{r_n}{w_n \!-\! r_0}\right]  \;,
\label{itqsm}
\end{equation}
This functions even for negative $a$, converging, in the present case,
after about 630 steps when $a=-0.3$.  For $a \leq -0.4$, however, the
scheme does not converge.  On the other hand, in the first example 
[i.e., rates given by Eq. (\ref{rtpa})], using $a<0$ does not offer 
any advantage over $a=0$.  

We conclude from these and other examples with large 
numbers of states \cite{rdunp},
that $a \simeq 0$ generally yields rapid convergence, but that
some amount of experimenting (for example, with negative $a$ values),
may reduce computation time further, in particular cases.
But even without such optimization, the iterative schemes of
Eqs. (\ref{its3}) and (\ref{itqss}) yield economies in computation time of two or
more orders of magnitude, compared with conventional integration methods.
As the number of variables and/or the disparity in the magnitudes
of the transition rates increases, direct iteration and the FI scheme become
ever more valuable.
\vspace{1em}

\noindent {\bf Acknowledgment}

I thank Dani ben-Avraham for helpful comments.
This work was supported by CNPq, Brazil.

\newpage

\noindent FIGURE CAPTIONS
\vspace{1em}

\noindent FIG. 1. Mean population $\langle n \rangle$ versus time in the coagulation
process,  Eq. (\ref{rtpa}), with $\lambda = 100$, using Runge-Kutta integration (solid line),
the second-order FI scheme with a fixed time step (circles), and
with a variable time step, as described in the text ($+$).
\vspace{1em}

\noindent FIG. 2. Stationary probability distribution in the coagulation process
with $\lambda = 400$.  Solid line: RKS; $+$: direct iteration of Eq. (\ref{its3}).
\vspace{1em}

\noindent FIG. 3. Quasi-stationary probability distribution in the generalized
Malthus-Verhulst process, Eq. (\ref{gmvp}), with
$\lambda = 3$, $\mu = 0.1$, and $\gamma = 1$.  
Solid line: RKS; $+$: direct iteration of Eq. (\ref{itqss}).
\vspace{1em}

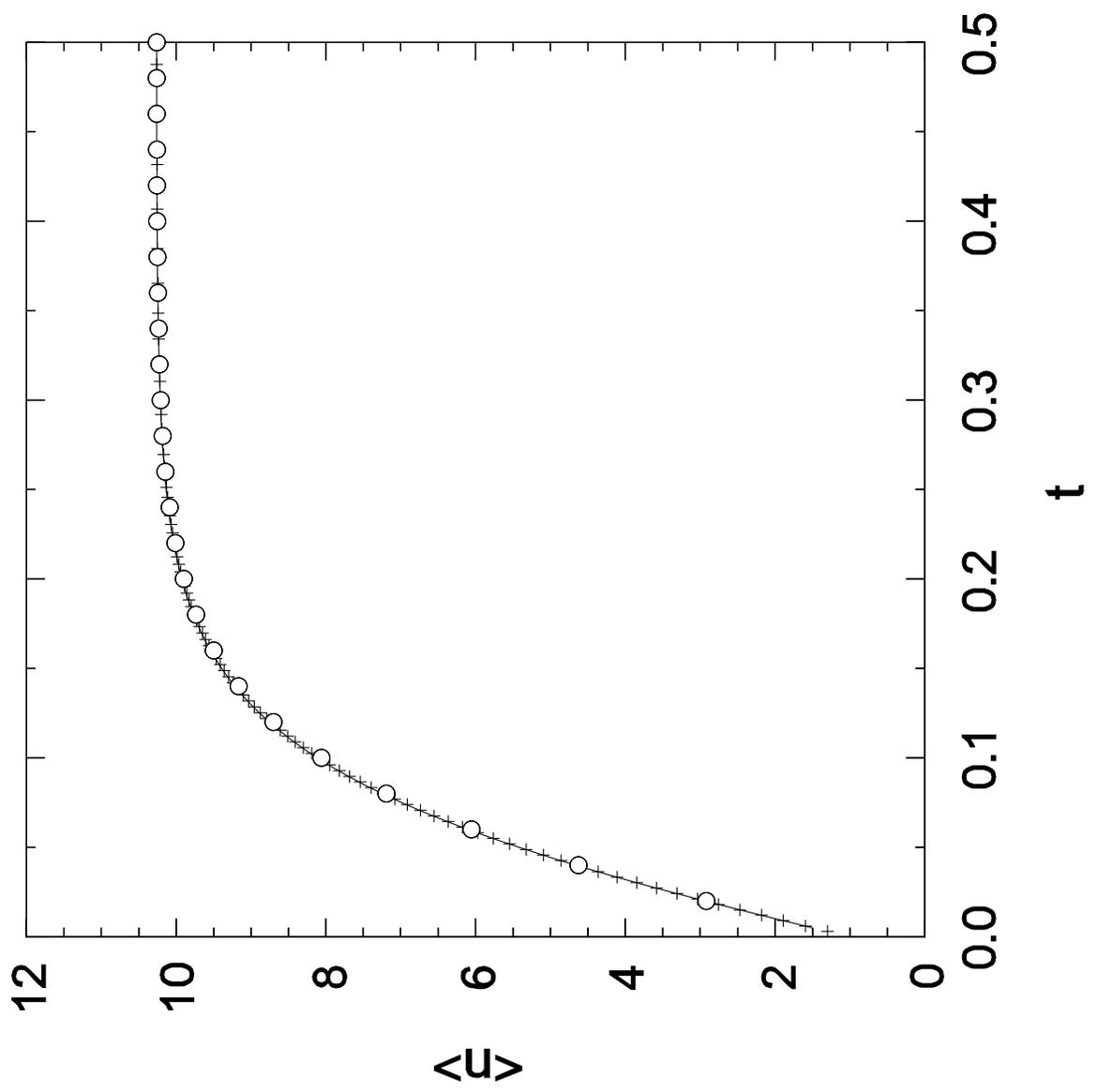
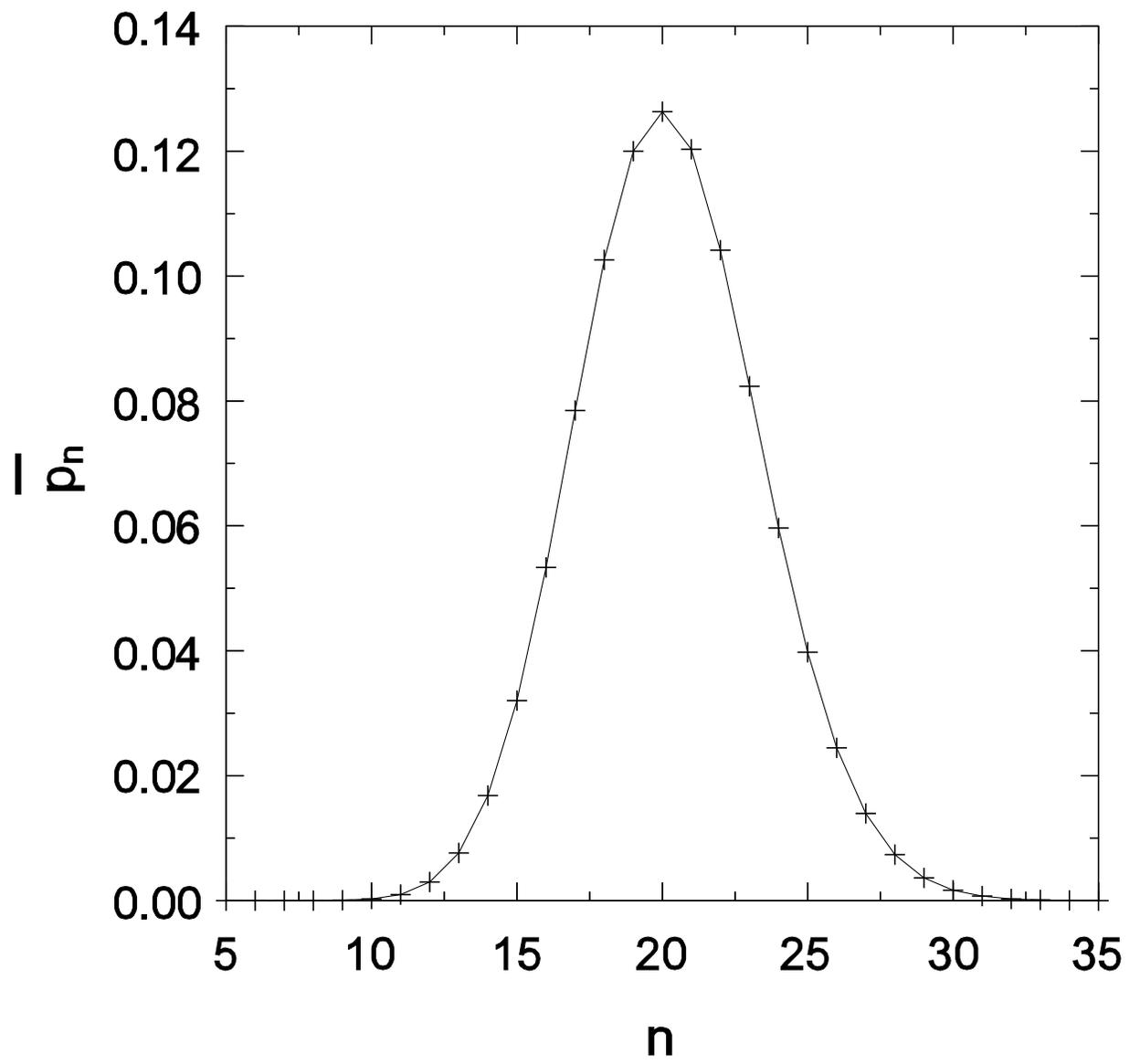
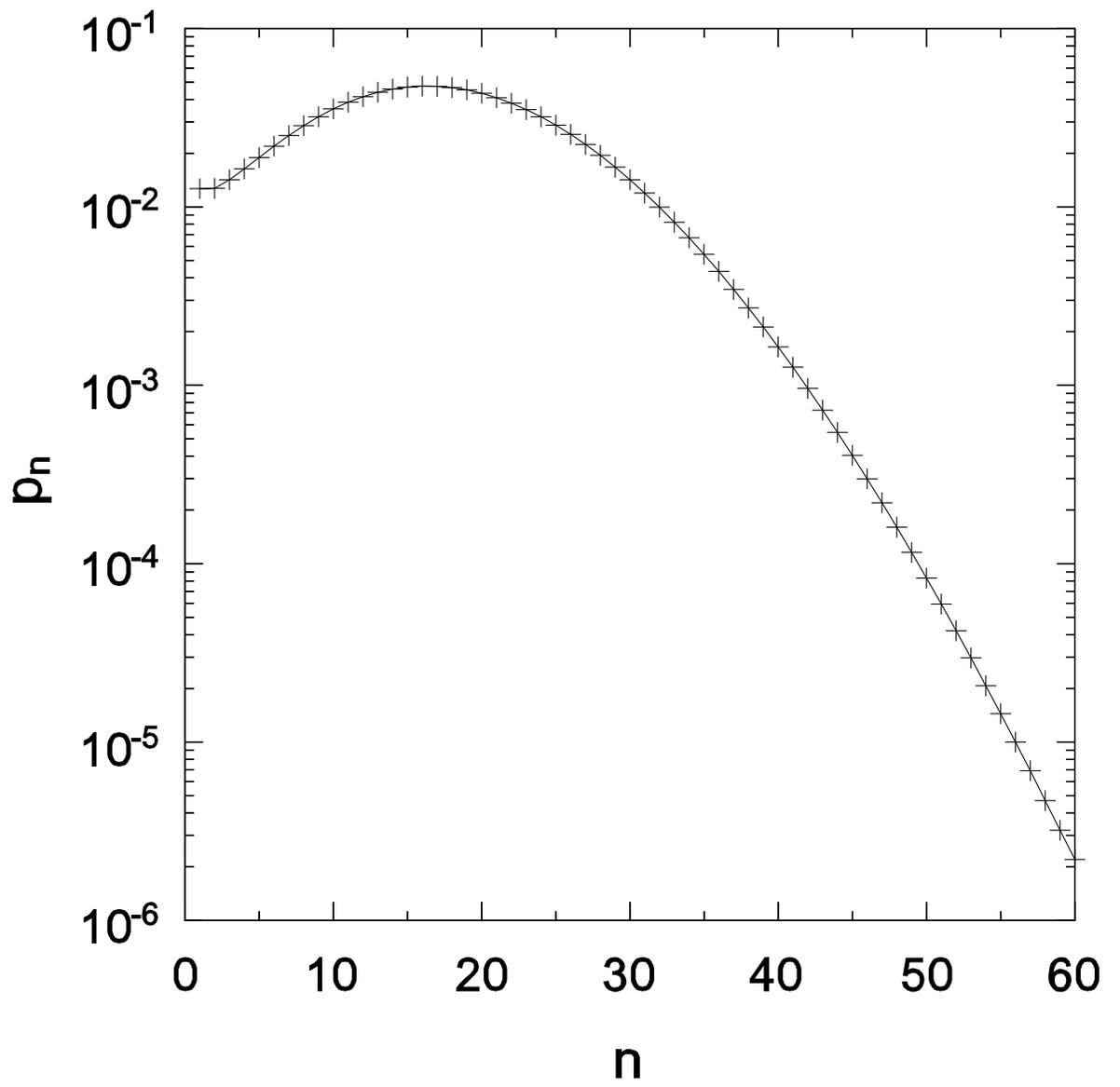
\end{document}